\documentclass[12pt]{article}

\usepackage{scicite}
\usepackage[usenames,dvipsnames]{xcolor}

\usepackage{graphicx}

\usepackage{times}

\usepackage{graphicx}
\usepackage{amsmath}
\usepackage{amssymb}
\usepackage{mathrsfs}
\usepackage{mathalfa}
\usepackage{changes}
\usepackage{amsfonts}
\usepackage{textcomp}

\definechangesauthor[name=Richard Hess, color=cyan]{rh}

\DeclareUnicodeCharacter{00B5}{\textmu}
\DeclareUnicodeCharacter{00B0}{\textdegree}

\newenvironment{sciabstract}{%
\begin{quote} \bf}
{\end{quote}}

\title{Gate-tunable topological superconductivity in a supramolecular electron spin lattice} 

\author
{ Rémy Pawlak,$^{1\ast\dagger}$ Jung-Ching Liu,$^{1\dagger}$ Chao Li,$^{1\dagger}$ Richard Hess,$^{1\dagger}$ Hongyan Chen,$^{2}$ \\
Carl Drechsel,$^{1}$, Ping Zhou,$^{3}$ Robert Häner,$^{3}$ Ulrich Aschauer,$^{3,4}$\\
Thilo Glatzel,$^{1}$ Silvio Decurtins,$^{3}$  Daniel Loss,$^{1}$ Jelena Klinovaja,$^{1}$\\
 Shi-Xia Liu,$^{3\ast}$ Wulf Wulfhekel,$^{2}$ \& Ernst Meyer$^{1}$\\
\\
\normalsize{$^{1}$Department of Physics, University of Basel, Klingelbergstrasse 82,}\\
\normalsize{ 4056 Basel, Switzerland}\\
\normalsize{$^{2}$Physikalisches Institut, Karlsruhe Institute of Technology,}\\
\normalsize{Wolfgang-Gaede-Str. 1, 76131 Karlsruhe, Germany}\\
\normalsize{$^{3}$Department of Chemistry, Biochemistry and Pharmaceutical Sciences,}\\
\normalsize{University of Bern, Freiestrasse 3, 3012 Bern,
Switzerland}\\
\normalsize{$^{4}$Department of Chemistry and Physics of Materials, University of Salzburg,}\\
\normalsize{Jakob-Haringer-Strasse 2A, 5020 Salzburg, Austria}\\
\\
\normalsize{$^\dagger$These authors equally contributed;}\\
\normalsize{$^\ast$To whom correspondence should be addressed;}\\
\normalsize{E-mails:  remy.pawlak@unibas.ch, shi-xia.liu@unibe.ch}
}

\date{}

\begin{document}

\baselineskip24pt
\maketitle 
\renewcommand{\figurename}{{\bf Fig.}}

\begin{sciabstract}
Topological superconductivity emerges in chains or arrays of magnetic atoms coupled to a superconductor. However, the external controllability of such systems with gate voltages is crucial for their future implementation in a topological quantum computer. Here we showcase the supramolecular assembly of radical molecules on Pb(111), whose discharge is controlled by the tip of a scanning tunneling microscope. Charged molecules carry a spin-1/2 state, as confirmed by observing Yu-Shiba-Rusinov in-gap states by tunneling spectroscopy at millikelvin temperature. Low energy modes are localized at island boundaries with an exponential decay towards the interior, whose spectral signature is consistent with Majorana modes protected by mirror symmetry. Our results open up a vast playground for the synthesis of gate-tunable organic topological superconductors.

\end{sciabstract}

\paragraph*{One sentence summary:}
Radical molecules placed on superconducting Pb(111) represent a novel platform for topological superconductivity


\cleardoublepage

\section*{Introduction}
Majorana zero modes (MZM) in topological superconductors obey non-Abelian statistics and are considered as the most promising building blocks for constructing topological qubits \cite{Nayak2008}. Topological superconductivity (TS) can be obtained in hybrid systems by coupling semiconducting nanowires with strong spin-orbit interaction \cite{Mourik2012,Lutchyn2018}, ferromagnetic atomic chains\cite{NadjPerge2014,Ruby2015,Pawlak2016,Feldman2017,Kim2018,Schneider2022}, or magnetic islands \cite{Menard2017,PalacioMorales2019,Wang2020,Kezilebieke2020} to a $s$-wave superconductor. One signature of MZMs is a zero-energy conductance peak, measured with a scanning tunneling microscope (STM) or in transport experiments. This zero-energy conductance peak appears close to the zero-dimensional boundary separating the trivial from the topological region. Higher dimensional boundaries  instead, lead to the formation of propagating edge states, namely chiral Majorana modes. Since local disorder can close the gap of non-trivial phase  by severely affecting the proximitized superconducting states\cite{Reeg2018,Prada2020,Hess2022}, assigning Majorana modes requires a fundamental understanding of the system parameters with high spectral resolution down to the atomic level\cite{Schneider2020,Ding2021,Kuester2022,Liebhaber2022}.\\

\noindent
The ability of STM to create and probe lattices with single-atom precision using manipulation techniques has offered unique opportunities for realizing designer quantum materials at the atomic scale \cite{Khajetoorians2019}. With the spectral resolution of scanning tunneling spectroscopy (STS), the detection of Yu-Shiba-Rusinov (YSR) states arising from magnetic moments on a superconductor \cite{Yu1965,Shiba1968,Rusinov1969} has revealed how important surface coordination\cite{Friedrich2021}, interatomic coupling \cite{Schneider2020,Ding2021,Kuester2022,Liebhaber2022}, or magnetic anisotropy\cite{Hatter2015} are to the formation of a complex topological phase  diagram. Moreover, the experimental analysis of these YSR bands in atomic structures has revealed the emergence of topological non-trivial phases by probing localized zero-bias peaks consistent with their topological origin, and thus attributed to MZMs  \cite{NadjPerge2014,Ruby2015,Pawlak2016,Feldman2017,Kim2018,Schneider2022,Menard2017,PalacioMorales2019,Wang2020,Kezilebieke2020}. \\

\noindent
Beyond densely-packed atomic structures governed by nearest neighbor exchange interaction \cite{NadjPerge2013}, dilute spin chains or two-dimensional "Shiba" lattices are also an exciting platform for the emergence of topological superconductivity \cite{Klinovaja2013,Pientka2013,Braunecker2013,Hoffman2015,Roentynen2015,Li2016,Rachel2017,Pawlak2019,Jaeck2021,Soldini2023}.  By increasing the spacing $a$ between magnetic impurities while keeping it smaller than the coherence length of the superconductor $\xi$, the YSR band formation remains possible by the in-gap state hybridization over a large number of nearest neighbors mediated by Ruderman--Kittel--Kasuya--Yosida interaction \cite{Klinovaja2013}. In such a regime (i.e. $k_{\rm F}a$ $\gg$ 1 with  $k_{\rm F}$ being the Fermi wave-vector), two-dimensional ferromagnetic Shiba lattices are predicted to exhibit a rich phase diagram with a large number of phases with high Chern numbers $\mathcal{C}$ \cite{Roentynen2015,Li2016,Rachel2017}, where chiral MZMs are localized at edges of the island with an exponential decay towards the island's interior. Recently, topological phases protected by spatial symmetries have been proposed to occur in a rich variety of topological crystalline superconductors\cite{Schnyder2015,Schindler2018,Geier2018}. Using STM, the first attempt to build and to probe such atomic lattices showed interesting signatures of edge modes consistent with a mirror-symmetry-protected topological superconductor \cite{Soldini2023}. \\

\noindent
While the physics of YSR states and edge states can be addressed by tunneling spectroscopy, the control over the chemical potential near these artificial structures, an essential prerequisite for future applications to tune the system with external gate voltages from trivial to topological, remains an open issue. Inspired by recent works on the charge-state control of organic molecules using the electric field of an STM tip \cite{FernandezTorrente2012,Kocic2015,Li2023}, our work explores the experimental realization of a two-dimensional spin lattice using the supramolecular assembly of gate-tunable radical molecules on superconducting Pb(111). This system could not only serve as a unique starting point for investigating the interplay of a prototypical array of electron spins with a superconductor but also provides a general playground for discovering  crystalline topological superconductivity in metal-free supramolecular network-superconductor hybrids.

\section*{Results and Discussion}
\paragraph*{\textit{Supramolecular assembly of TBTAP molecules on superconducting  Pb(111)}.}
As precursor, we used the 4,5,9,10-tetrabromo-1,3,6,8-tetraazapyrene (TBTAP) molecule (Fig.~1A) consisting of an electron acceptor tetraazapyrene backbone equipped with four peripheral bromine atoms \cite{Zhou2021}. We recently showed that TBTAP$^{\bullet-}$ radicals with a 1/2 spin state retain a single electron on Ag(111) without using a decoupling layer \cite{Li2023}. To obtain large supramolecular domains of more than 100 nanometers in diameter, TBTAP molecules   were sublimed in ultra high vacuum on a Pb(111) substrate kept at about 200 K \cite{Methods} (Fig.~1B). A densely packed rectangular network of lattice parameter $a_1$ = 12.3~\AA{} and $b_1$ = 17.2~\AA{} (arrows in Fig.~1C), is observed by STM as alternating dark and bright rows. The corresponding image obtained by atomic force microscopy (AFM) (Fig.~1D) shows each Br atom bound to the TAP backbone as bright protrusion allowing us to assign the exact molecule position in the array (see models in Fig.~1D). Similar to STM imaging, two AFM contrasts are observed as a function of the considered rows denoted in the following as charged ({\it c}) (dashed line)  and neutral ({\it n}) (dotted line), which will be discussed later.\\

\noindent
Using density functional theory (DFT) \cite{Methods}, we relaxed the TBTAP network on Pb(111) (Fig.~1E, supplementary text and fig.~S1). The assembly is in registry with the Pb(111) surface in agreement with the experimental data ($a_1$ = 12.1~\AA~  and $b_1$ = 17.5~\AA~, fig. S1). Molecules are stabilized by a combination of halogen bonds between Br atoms (C--Br...Br--C) and TAP units (C--N...Br--C). 
Dashed and dotted lines correspond to rows of charged ({\it c}) and neutral ({\it n}) molecules, respectively. 
Molecules lie flat in a plane 3.4~\AA~above the surface, suggesting that the variation of STM/AFM contrasts between neighboring rows is due to the coexistence of two molecule charge states in the network \cite{Kocic2015}, rather than a difference in relative height (see fig. S1). 

\paragraph*{\textit{Electrical control of molecule's charge state in the assembly}.}
To confirm this, we compared d{\it I}/d{\it V} point-spectra  of TBTAP molecules located in {\it c} and {\it n} rows, respectively (Fig.~2B). Molecule {\it c} (blue spectra) shows a strong resonance {\it D} at $V_{\rm D}$ $\approx$ 1 V, which is absent for molecule {\it n} (red spectra). The {\it D} resonance is assigned to a charge-state transition induced by the local electric field of the tip from the anionic TBTAP$^{\bullet-}$ molecule to its neutral TBTAP$^{0}$ counterpart\cite{FernandezTorrente2012,Kocic2015,Li2023}. Without gating, radical TBTAP$^{\bullet-}$ molecules are obtained by the transfer of one electron from the surface to the lowest unoccupied molecular orbitals (LUMO) \cite{FernandezTorrente2012,Li2023}, leading to the LUMO splitting into a singly-unoccupied molecular orbital (SUMO) and a singly-occupied molecular orbital (SOMO) (see supplementary text, figs. S2-S4). Charging events expected as a dip in d{\it I}/d{\it V} spectra for negative voltages were not observed for both type of molecules.\\

\noindent
Figure~2C shows a constant-height d{\it I}/d{\it V} map acquired at the threshold voltage $V_{\rm D}$. Rings/dots of high conductance centered to  molecules {\it c} are the hallmark of a successful discharge, which also indicates the spatial position of the electron in TBTAP$^{\bullet-}$ molecules prior to its removal. Using the double-junction tunneling barrier (DJTB) model~\cite{FernandezTorrente2012,Kocic2015,Li2023}, the efficiency with which the tip locally discharges nearby molecules is characterized by the lever arm $\mathcal{L}$, which  at first approximation linearly depends on the tip-sample voltage $V_{\rm S}$ and its position with respect to the molecule (supplementary text, figs. S5-S6). 
Figure~\ref{Fig2}D shows a d{\it I}/d{\it V} cross-section acquired across {\it n-c-n} rows (plain line in Fig.~2A). Discharging rings are absent along {\it n} rows since no charge can be extracted from neutral TBTAP$^{0}$. The discharge parabola is centered to the TBTAP$^{\bullet-}$ with its bottom ($\approx$ 0.9 V) corresponding to the resonance. Due to  the linear voltage-dependency of $\mathcal{L}$  the parabola branches expand with increasing $V_{\rm S}$, thus reflecting the increase in size of rings in d{\it I}/d{\it V} maps with increasing voltages (Figs.~\ref{Fig2}F-I, see supplementary text).\\

\noindent
At fixed  $V_{\rm S} \geq V_{\rm D}$, discharging rings form a superlattice of parameter $a_2$ = 20~\AA~and $b_2$ = 39.7~\AA~and rotated by 30° with respect to the molecular lattice observed in d{\it I}/d{\it V} mapping (Fig.~\ref{Fig2}C). Their diameters vary between neighboring TBTAP$^{\bullet-}$ as the result of a local modulation of the resonance, estimated to $\approx$ 150 mV by comparing the bottom of each parabola (dashed line in Fig.~\ref{Fig2}E). For $V_{\rm S}$ $\geq$ 1.1 V and when the tip is located between two neighboring molecules, the parabolas start to merge promoting the removal of two electrons $2e$ from the neighboring molecules (region $e$ $\geq$ 1). Accordingly, increasing $V_{\rm S}$ in a series of spatial d{\it I}/d{\it V} maps leads to ring expansion (Fig.~2G) followed by their coalescence (Fig.~2H). In contrast to a simple superposition of rings expected for non-interacting quantum dots, their fusion as observed in Figs.~2H-I indicates a cascade discharge along {\it c} rows and thus a manifestation of the electron correlation in the supramolecular assembly \cite{Li2021} (see supplementary text, fig. S6).\\

\paragraph*{\textit{Yu-Shiba-Rusinov bound states of radical molecules}.}
Radical TBTAP$^{\bullet-}$ on Pb(111) feature a S = 1/2 ground state with a strong spin-polarization according to DFT calculations (Fig.~3A). We probed YSR bound states of radical TBTAP$^{\bullet-}$ in the middle of a molecular island using tunneling spectroscopy with a metallic tip at $T$ = 35 mK (Fig.~3B) \cite{Methods}. Figure~\ref{Fig3}C compares d{\it I}/d{\it V} spectra of three representative TBTAP$^{\bullet-}$ molecules marked in Fig.~\ref{Fig3}B (black spectra, Fig.~\ref{Fig3}C) with that of Pb(111) (blue) and a neutral TBTAP$^{0}$ (red). For the last two, a hard gap centered to $E_{\rm F}$ and framed by the two coherence peaks of Pb(111) \cite{Ruby2015a} is systematically measured without in-gap states. Each TBTAP$^{\bullet-}$ spectrum additionally shows one pair of YSR states at energies $\varepsilon _{\alpha}$ = $\pm$ 460 µeV, $\varepsilon _{\beta}$ = $\pm$ 720 µeV and $\varepsilon _{\gamma}$ = $\pm$ 940 µeV (dotted lines), resulting from the spin-1/2 nature of radical TBTAP$^{\bullet-}$.  By applying an out-of-plane magnetic field of 0.5 T, we also quenched the superconductivity state to probe the Kondo resonance (figs. S8) and estimated its Kondo temperature $T_{\rm K}$ to be 10.3 K (figs. S7)~\cite{Li2023}. Electron-like and hole-like wave-functions of TBTAP$^{\bullet-}$ were also probed by d$I$/d$V$ mapping at the $\varepsilon_{\alpha}^{\pm}$, $\varepsilon_{\beta}^{\pm}$ and $\varepsilon_{\gamma}^{\pm}$ energies (Figs.~\ref{Fig3}E-G), respectively. The typical donut-shape is similar to the spin density map (Fig.~\ref{Fig3}A), while their energies  depends on the molecule positions in the assembly. We also infer that the shift of the YSR states to higher energies as compared to that of the isolated molecule (fig. S9) and their  spatial distribution points to a coupling of the quasi-particle excitations within the supramolecular network \cite{Liebhaber2022}. Figure~\ref{Fig3}G shows a d{\it I}/d{\it V} cross-section across the island (red arrow of Fig.~\ref{Fig3}D), where white dotted lines refer to the $\varepsilon_{\alpha,\beta,\gamma}^{\pm}$ YSR energies. Broader resonances near $E_{\it F}$ coexist with the YSR peaks as marked by black arrows in Figs.~\ref{Fig3}C and G. These low-energy modes (LEM) are systematically observed with the highest magnitude for molecules aligned along the white dashed lines marked in the zero-energy d{\it I}/d{\it V} map of Fig.~\ref{Fig3}D.\\ %

\paragraph*{\textit{Spectral signature and localization length of low-energy modes near an island edge}.}
The intrinsic electron-hole symmetry of zero-energy modes, imposed by the Bogoliubov quasi-particle character, can be probed by tunneling spectroscopy using superconducting STM tips ($\Delta_{\rm T}$ = 1.35 meV is the superconducting pairing energy of the tip). Experimentally, a zero-energy peak appears in d{\it I}/d{\it V} spectra as a pair of peaks of equal amplitude shifted from zero to the finite voltages {\it eV} = $\pm$ $\Delta_{\rm T}$, while the superconducting edge is observed at $\pm$ ($\Delta_{\rm T}$ + $\Delta_{\rm S}$) = $\pm$ 2.7 meV ($\Delta_{\rm S}$ = 1.35 meV is the superconducting gap of the substrate).  Using bulk Pb tips at $T$ = 1 K (Fig.~\ref{Fig4}) \cite{Methods}, we confirmed the presence of YSR in-gap states at the TBTAP$^{\bullet-}$  locations by d{\it I}/d{\it V} point-spectra (Fig.~\ref{Fig4}B) acquired along seven TBTAP$^{\bullet-}$ molecules of a {\it c} row (Fig.~\ref{Fig4}A, fig. S12). Due to the larger thermal broadening of $\approx$ 90-100 µeV at 1K, the accurate assignment of the $\varepsilon_{\alpha, \beta, \gamma}$  energies is less evident than that of the millikelvin measurements since these peaks merge into a single resonance found at {\it eV} = $\pm (\Delta_{\rm T}$ +  $\varepsilon _{\alpha, \beta, \gamma}$) $\approx$ $\pm$ 2.1 meV. Note also that the YSR states are always accompanied by a broader resonance near zero-energy (i.e. $\pm$ $\Delta_{T}$) which is the fingerprint of the low-energy modes using superconducting tips.  \\

\noindent
The constant-height d{\it I}/d{\it V} maps of Figs.~\ref{Fig4}D and E compare the spatial distribution near the edge of an island of the hole-like wavefunctions extracted at $\varepsilon^+$ with the LEM wavefunctions at +$\Delta_{T}$. While the DOS at the YSR energy is homogeneous along the {\it c} rows, the LEM lines emerge from the ferromagnetic edge schematized in the model of Fig.~\ref{Fig4}F. They propagate along the direction rotated by 60° with respect to the edge corresponding to a ferromagnetic direction of the spin structure.  
Figure~\ref{Fig4}G shows a d{\it I}/d{\it V}(V,X) cross-section acquired along one LEM line marked by a blue dashed line in Fig.~\ref{Fig4}E. All sub-gap excitations now appear at zero energy with equal amplitudes between electron-like and hole-like regions (Fig.~\ref{Fig4}H). This observation, in stark contrast with the strong intrinsic electron-hole asymmetry of the YSR resonances (Fig.~\ref{Fig4}B), underlines the zero-energy character of these edge modes.\\

\noindent
We next characterize the LEM localization length (Fig.~\ref{Fig4}I) by comparing d{\it I}/d{\it V}(X) profiles along the LEM line (blue) with that obtained at the YSR  energy (gray) (see also supplementary text, fig. S13). In contrast to the continuous DOS at $\varepsilon^{+}$, the LEM wavefunction has a maximum amplitude at the border of the island ($X$ = 0)  and decays towards the interior but without completely vanishing. In Fig.~\ref{Fig4}I, we estimated the experimental decay of the edge mode by fitting its envelope (dashed line) with a function $f(x)$  composed of two exponents representing the short $\xi_1$ = 3 nm and the long $\xi_2$ = 110 nm localization length\cite{Pawlak2016}. We explain it by considering the TBTAP$^{\bullet-}$ network as a lattice of spin-1/2 impurities with long-range YSR overlap coupled to a superconductor, as reported in References \cite{Roentynen2015,Li2016}. \\

\begin{figure}
	\centering
		\includegraphics[width=0.45\textwidth]{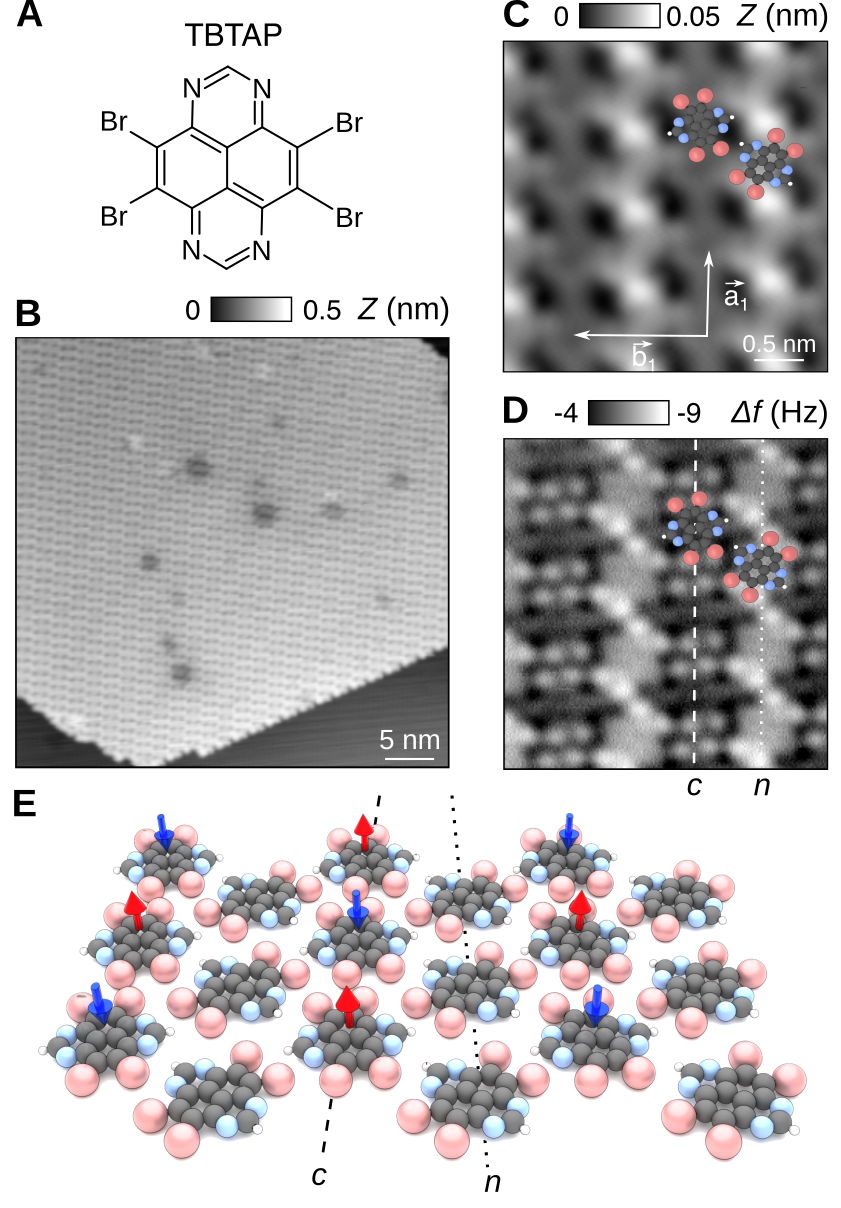}
		\caption{{\bf Supramolecular assembly of radical TBTAP$^{\bullet-}$ molecules on Pb(111).}
		(\textbf{A}), Chemical structure of the 4,5,9,10-tetrabromo-1,3,6,8-tetraazapyrene (TBTAP) molecule.
		(\textbf{B}), STM overview of the supramolecular assembly on Pb(111) ($V_{\rm s}$ = -0.5 V, $I_{\rm t}$ = 1 pA).
		(\textbf{C}), Close-up STM image of the molecular lattice ($V_{\rm s}$ = 0.8 V, $I_{\rm t}$ = 0.8 pA). 
		The unit cell is marked with arrows.
		(\textbf{D}), Corresponding AFM image acquired with Br-terminated tip.  The molecular lattice consists of rows of charged ($\textit{c}$) and neutral ($\textit{n}$) TBTAP molecules.  Red, blue, white and black colors of the ball-and-stick models refer to bromide, nitrogen, hydrogen and carbon atoms. (\textbf{E}), Perspective view of the TBTAP assembly on Pb(111) optimized by DFT calculations (fig. S1).  Dashed and dotted lines correspond to rows of charged ({\it c}) and neutral ({\it n}) molecules, respectively. Red and blue arrows represent the spin lattice of the TBTAP assembly. \label{Fig1}}
\end{figure}

\begin{figure}
	\centering
	\includegraphics[width=0.95\textwidth]{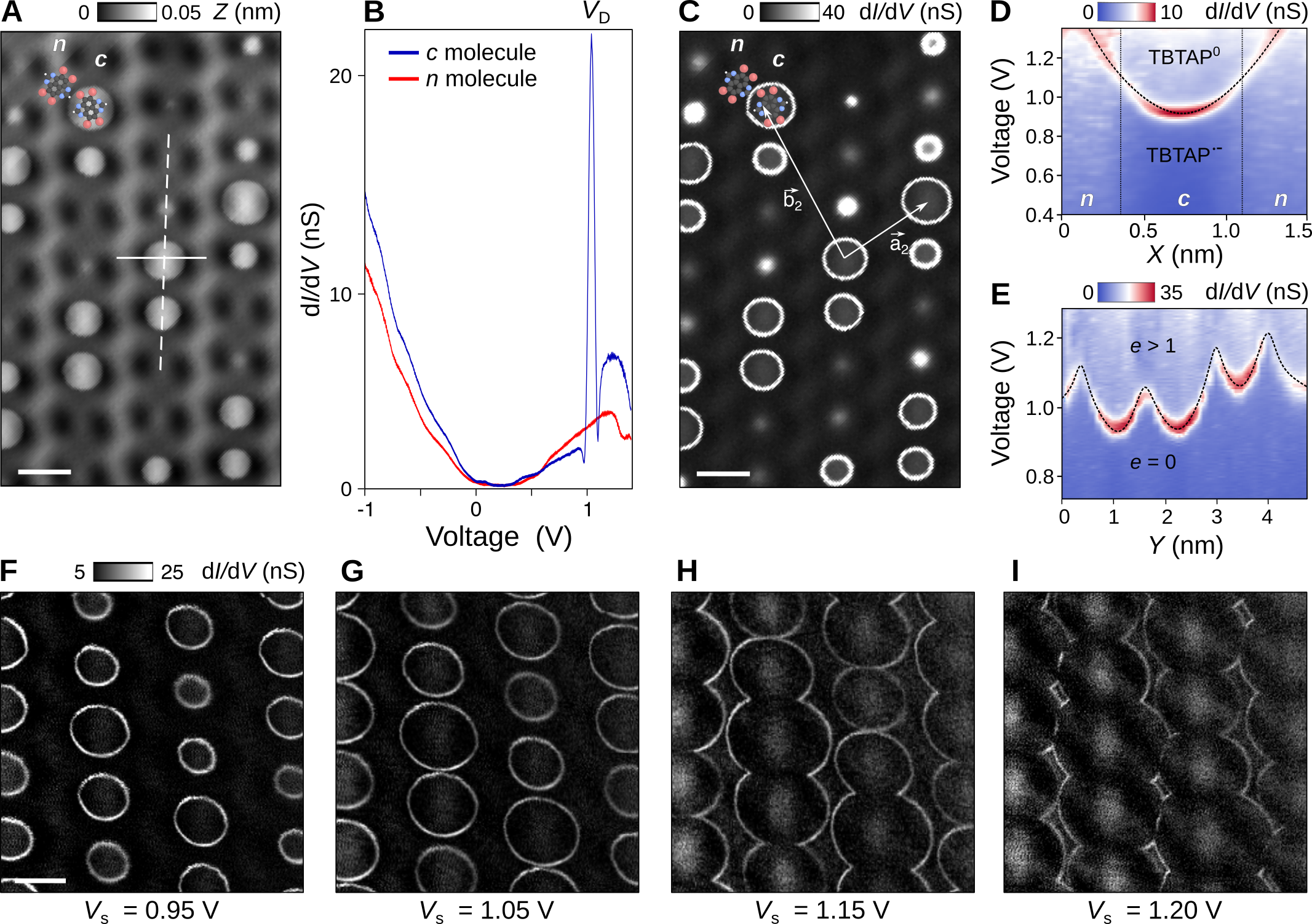}
	\caption{{\bf Electron superlattice of the supramolecular assembly.}
	(\textbf{A}), STM image of the molecular assembly ($V_{\rm s}$ = 0.8 V, $I_{\rm t}$ = 50 pA). 
(\textbf{B}), Single-point d{\it I}/d{\it V} spectra of {\it c} (blue) and {\it n} (red) molecules. The {\it D} resonance peak corresponds to the tip-induced discharge event (Set-points : 100 pA, -0.8 V, lock-in parameters: $f$ = 611 Hz, $A_{\rm mod}$ = 20 mV). 
(\textbf{C}) d{\it I}/d{\it V} mapping showing discharge rings centered to each molecule of {\it c} rows.
(\textbf{D}), d{\it I}/d{\it V} cross-section across {\it n-c-n} rows (plain line in \textbf{A}) showing  the electron localization. 
(\textbf{E}), d{\it I}/d{\it V} cross-section across five TBTAP$^{\bullet-}$ molecules (dashed line in \textbf{A}). The dashed line corresponds to multiple discharge events ($e$ is the number of removed charges) induced by tip gating.
(\textbf{F-I}), Series of d{\it I}/d{\it V} mapping for increasing tip-sample voltage $V_{\rm S}$ showing the expansion of the ring diameter and cascade discharge for $V_{\rm s}$ $\geq$ 1.15 V. Scale bars are 1 nm.\label{Fig2}}
\end{figure}

\begin{figure}
	\centering
	\includegraphics[width=0.95\textwidth]{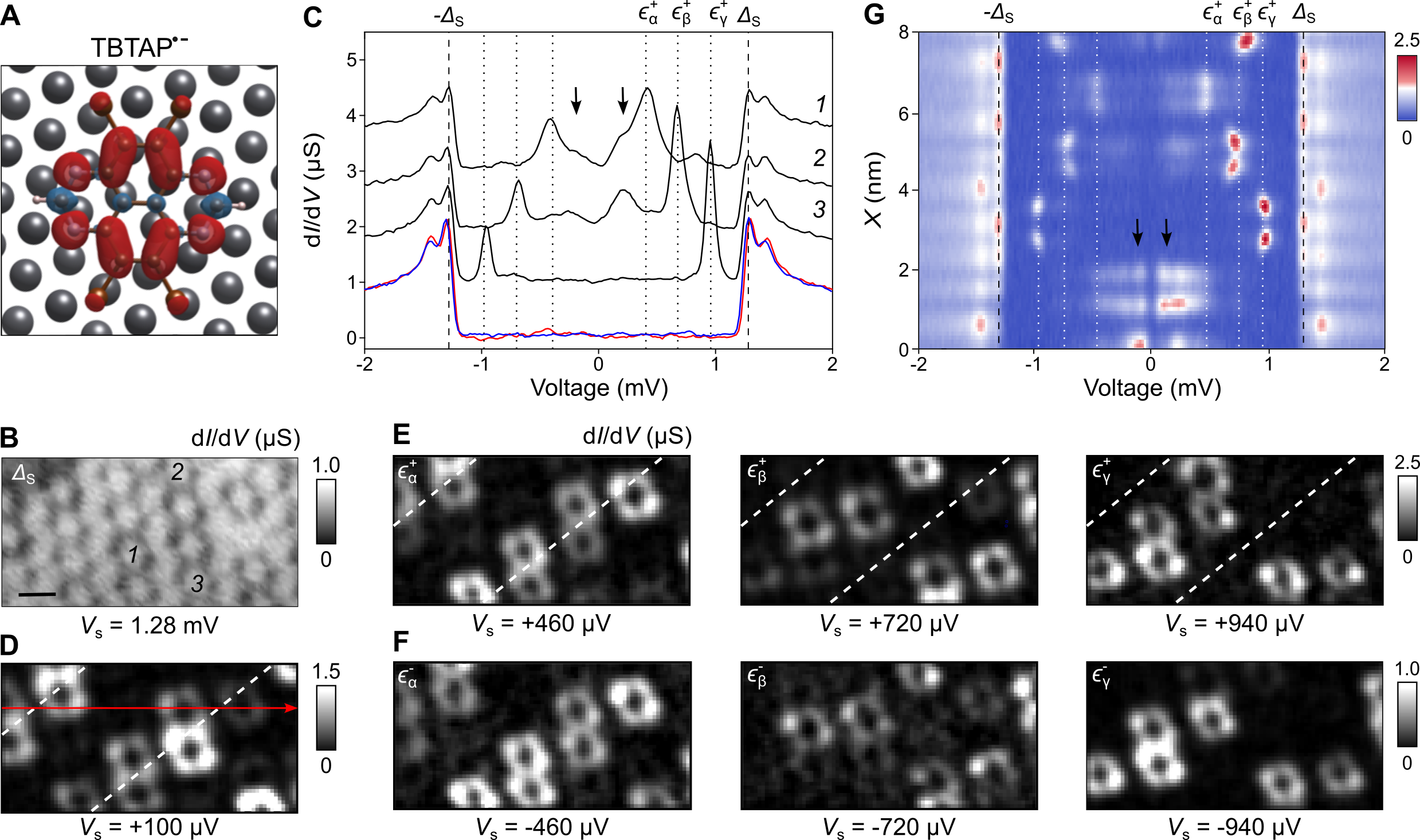}\\ 
	\caption{{\bf Yu-Shiba-Rusinov states of TBATP$^{.-}$ molecules.}
	(\textbf{A}), Spin density map of the TBTAP$^{\bullet-}$  molecule on Pb(111) calculated by DFT.
	(\textbf{B}) d{\it I}/d{\it V} map of the TBTAP$^{\bullet-}$ assembly acquired at $T$ = 35 mK with a metallic tip for $V_{\rm S}$ = 1.28 meV (scale bar is 1 nm, lock-in parameters: $f$ = 3.28 kHz, $A_{\rm mod}$ = 40 µV, tunneling parameters: $I_{\rm T}$ =  200 pA, $V_{\rm S}$ =  5 mV.).
(\textbf{C}), Exemplary d{\it I}/d{\it V} point-spectra of Pb(111) (red) and a neutral TBTAP$^{0}$ (blue). Black curves are three representative spectra of TBTAP$^{\bullet-}$ molecules marked in \textbf{B}. Spectra are shifted by 1 µS for clarity. Pairs of YSR resonances at $\varepsilon_{\alpha}$ = $\pm$ 460 µeV, $\varepsilon_{\beta}$ = $\pm$ 720 µeV and $\varepsilon_{\gamma}$ = $\pm$ 940 µeV are molecule-dependent along the $c$ row. The black arrows mark the low-energy modes near $E_{\rm F}$. 
(\textbf{D}), Near zero-energy d{\it I}/d{\it V} map ($\varepsilon$ = +100 µeV) showing the spatial distribution of the LEM lines (white dashed lines). 
(\textbf{E-F}) Series of d{\it I}/d{\it V} maps extracted at energies of $\pm$ $\varepsilon_{\alpha}$, $\pm$ $\varepsilon_{\beta}$ and $\pm$ $\varepsilon_{\gamma}$, respectively.
(\textbf{G}), d{\it I}/d{\it V}(V,Y) cross-section acquired along the red arrow of \textbf{D}. White dotted lines refer to $\pm$ $\varepsilon_{\alpha}$, $\pm$ $\varepsilon_{\beta}$ and $\pm$ $\varepsilon_{\gamma}$ energies. Black dashed lines show the superconducting edge $\pm$ $\Delta$ while black arrows mark the LEM.\label{Fig3}}
\end{figure}

\begin{figure}
	\centering
	\includegraphics[width=0.95\textwidth]{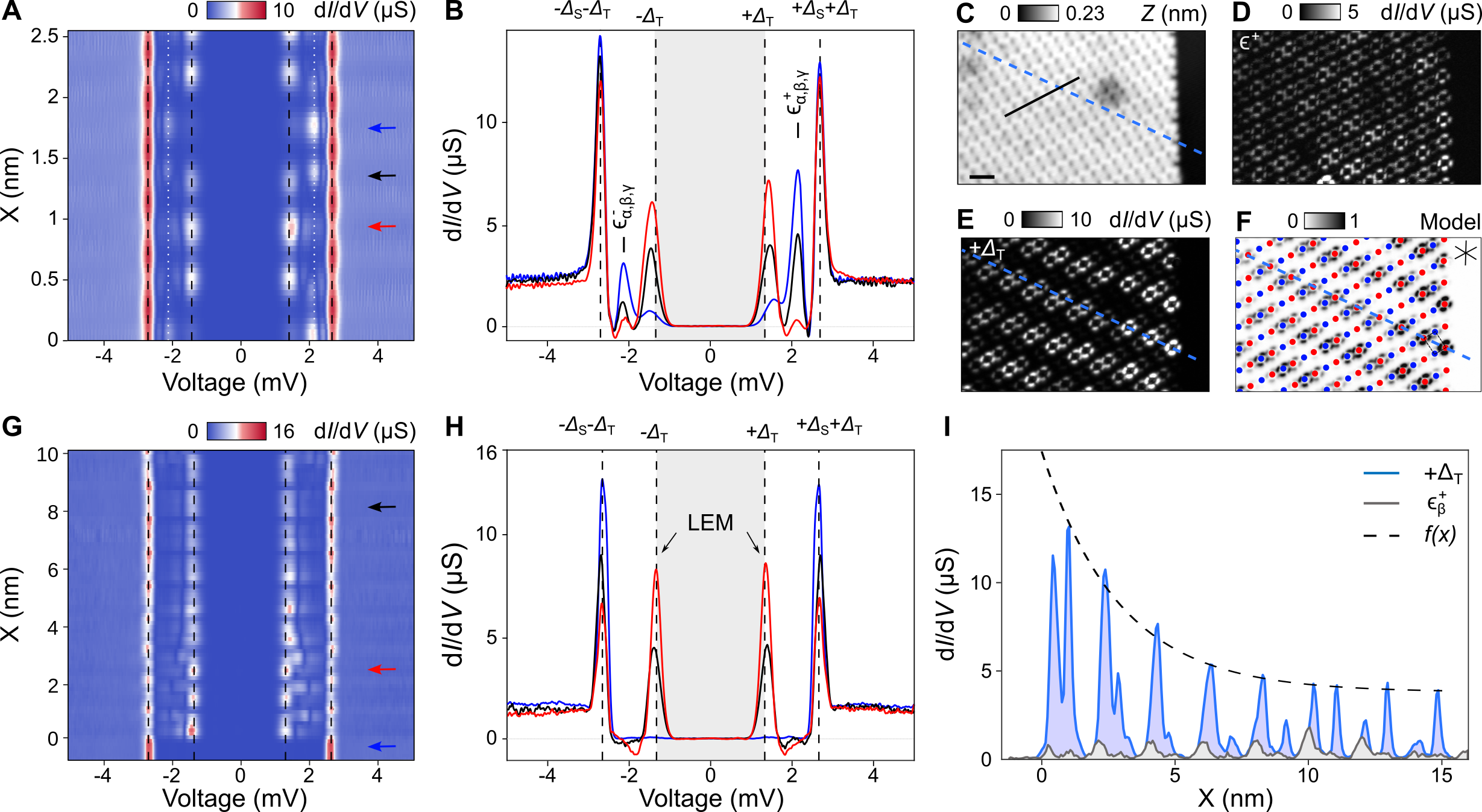}\\
	\caption{{\bf Tunneling spectroscopy with superconducting tips of the LEM near an edge .}
	(\textbf{A}), d{\it I}/d{\it V} cross-section acquired along seven molecules of a {\it c} row (black line in {\bf C}) (lock-in parameters: $f$ = 611 Hz, $A_{\rm mod}$ = 25 µV, tunneling parameters: $I_{\rm T}$ = 300 pA, $V_{\rm S}$ =  5mV). White dotted lines are the  $\varepsilon^{\pm}_{\alpha, \beta, \gamma}$ energies. Dashed lines correspond to $\pm$ ($\Delta_{\rm T}$ + $\Delta_{\rm S}$) and $\pm$ $\Delta_{\rm T}$, respectively.
		(\textbf{B}), d{\it I}/d{\it V} spectra extracted at the red, blue and black arrows of \textbf{A}. 
	(\textbf{C}), STM topographic image of and edge and (\textbf{D-E}), d{\it I}/d{\it V} maps acquired near the island edge of the LEM  wavefunction and the $\varepsilon^+_{\alpha, \beta, \gamma}$ energy.  Scale bar is 1 nm.
(\textbf{F}), Schematic of spin structure superimposed on top of the LEM wavefunction. Red and blue dots refer to spin up and spin down, respectively. LEM lines are rotated by 6° as compared to the $[1\bar{1}0]$ directions of Ag(111) (dark lines).
	(\textbf{G}), d{\it I}/d{\it V} cross-section across the edge taken along one LEM line (blue dashed line of {\it C}). 
	(\textbf{H}), d{\it I}/d{\it V} spectra at the blue, red and black arrows of {\bf G}.  The LEM appears as a pair of peaks of equal magnitude at $\pm \Delta_{\rm T}$.
	(\textbf{I}), Extracted decay length of the LEM wavefunction (blue) as compared to the $\varepsilon^+_{\alpha, \beta, \gamma}$ energy (gray) along the blue dashed line of {\it C}. The island border is set at $X$ = 0 nm. The dashed line is a fit of the edge mode envelope with two exponential functions that estimate a short ($\xi_1$ = 3 nm) and a long ($\xi_2$ = 110 nm) localization length.
	 \label{Fig4}}
\end{figure}

\begin{figure}
	\centering
	\includegraphics[width=0.65\textwidth]{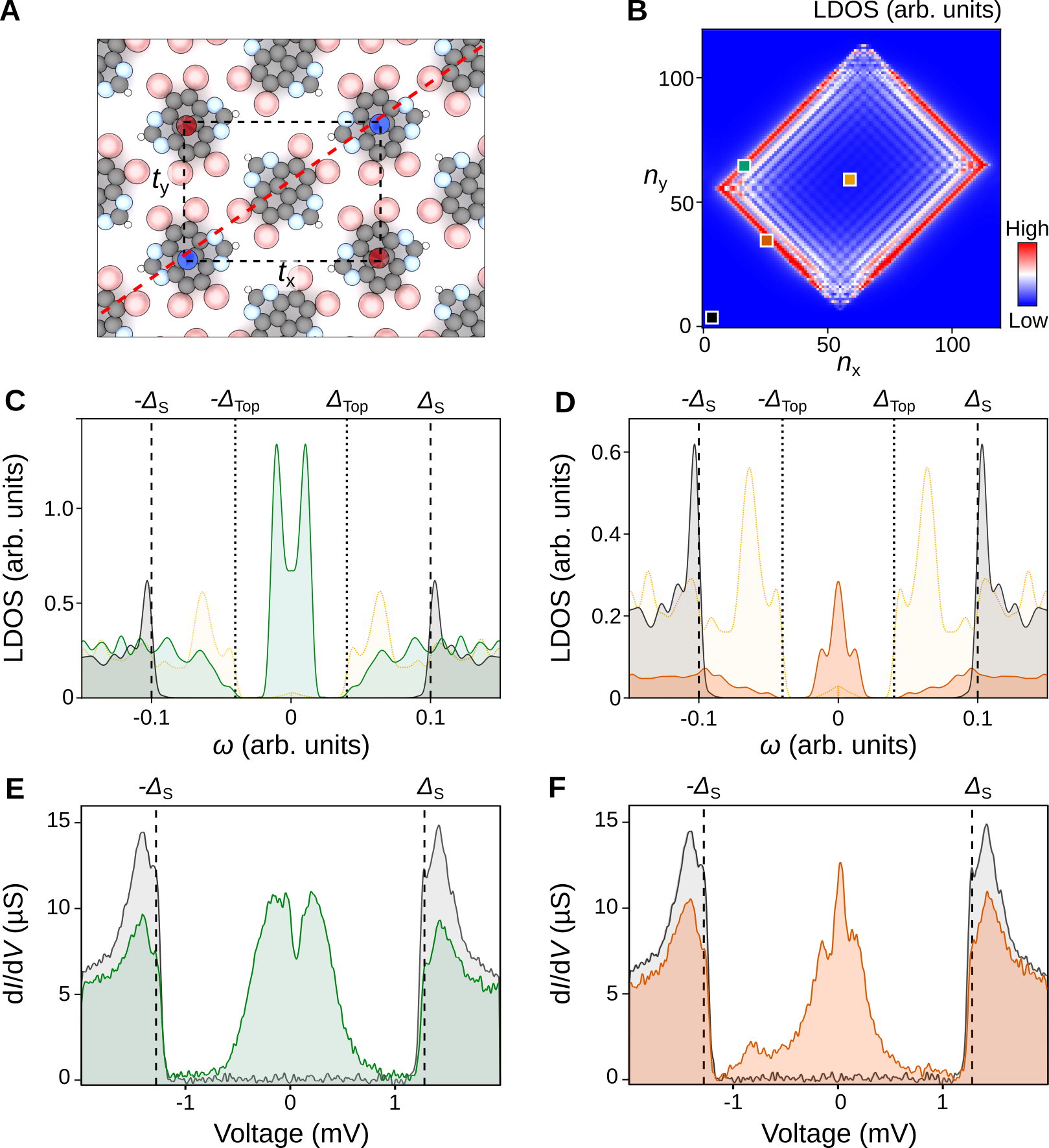}\\
	\caption{{\bf Comparison of LEM theoretical modeling with experimental data.}
	(\textbf{A}), Antiferromagnetic spin lattice (dashed) formed by the supramolecular assembly (red, spin up; blue, spin down). The molecular lattice imposes different hopping strengths $t_{\rm x}$ and $t_{\rm y}$ in $x$ and $y$ directions. Red dashed line is the assumed 45°-edge with respect to the molecular lattice.
	(\textbf{B}), Spatial LDOS map of the lattice at zero energy showing the emergence of edge modes. 
	(\textbf{C}-\textbf{D}), Theoretical LDOS($\omega$) spectra of the LEM extracted at edges (green and orange squares in \textbf{A}) as compared to the pristine Pb spectrum (black) and the spectrum at the center of the island (yellow). Dashed and dotted lines refer to the superconducting ($\pm$ $\Delta_{\rm S}$) and topological gaps ($\pm$ $\Delta_{\rm Top}$), respectively.
	(\textbf{E}-\textbf{F}), Experimental d{\it I}/d{\it V} spectra of the edge modes acquired at 50 mK with a metallic tip (lock-in parameters: $f$ = 3.28 kHz, $A_{\rm mod}$ = 10 µV, tunneling parameters: $I_{\rm T}$ =  200 pA,$V_{\rm S}$ =  5mV.).  \label{Fig5}}
\end{figure}

 
\paragraph*{\textit{Theoretical analysis}.}
To further rationalize our findings, we used a tight binding model on a rectangular spin lattice similar to one introduced by Soldini {\it et al.} \cite{Soldini2023} in order to describe the spatial-symmetry-protected topological order of an antiferromagnet-superconductor hybrid structure \cite{Methods}. As suggested by our tunneling measurements (fig. S10), we assume an antiferromagnetic ordering of the spin (schematized by red and blue arrows in Fig.~\ref{Fig5}A and  Fig. 1E), along the lattice imposed by the TBTAP molecular assembly (dashed line). Based on our STM observations (Fig.~\ref{Fig1}B and Fig.~\ref{Fig4}C), we construct a prototypical "Shiba" island mimicking the supramolecular network boundaries by considering only ferromagnetic edges along the [110] directions with respect to the molecular lattice (red line in Fig.~\ref{Fig5}A). Figure~\ref{Fig5}B shows the calculated zero-energy LDOS map of the system, which demonstrates the formation of edge modes in agreement with reference~\cite{Soldini2023}. Theoretical LDOS spectra are plotted in Figs.~\ref{Fig5}C and D for two edge positions marked by green and red squares in Fig.~\ref{Fig5}B, respectively. These edge modes have two typical spectral signatures consisting of either two peaks of equal amplitude split from zero energy (green in Fig.~\ref{Fig5}C) or three resonances centered to zero energy (red in Fig.~\ref{Fig5}C). They are both framed by a topological gap ($\pm \Delta_{\rm Top}$) extracted at the center of the island (yellow spectra in Figs.~\ref{Fig5}C-D) as well as the superconducting gap  at $\pm \Delta_{\rm S}$ (black spectra). Importantly, the 45°-edges of the antiferromagnetic island respect the underlying spatial symmetries, namely mirror symmetries, \cite{Soldini2023}, such that a gapped topological crystalline superconducting phase with topological edge states can form. The experimental spectroscopic signatures of the LEM (Figs.~\ref{Fig5}E and F), acquired near an edge at 50 mK with a metallic tip (see figs. S14 and S15), is in good agreement with the theoretical predictions of these topological edge modes.

\section*{\large Conclusion and outlook}
In conclusion, we demonstrate the formation of an extended array of electron spins on superconducting Pb(111) through the supramolecular assembly of organic radicals. Occupied by a single electron transferred from the substrate, radical molecules are in a spin-1/2 ground state confirmed by probing one pair of Yu-Shiba-Rusinov in-gap states in differential conductance spectra. In the two-dimensional supramolecular assembly, spectroscopic signatures of low energy modes (LEM) are observed in tunneling spectroscopy near edges of the island with a long decay towards the interior. Using both metallic and superconducting tips, we characterized the near zero-energy character of this resonance, its intrinsic particle-hole symmetry, the localization length as well as site-dependent spectral signatures. Altogether, these key features confirmed by theory are consistent with the emergence of topological non-trivial modes in an antiferromagnet-superconductor hybrid structure that can be assigned to Majorana modes \cite{Soldini2023}.\\

\noindent
Such a spatial-symmetry-protected topological superconductor has a complex phase diagram which crucially depends on the edge terminations of the system boundaries (fig. S10) as well as the lattice parameter $a$ (i.e. hopping parameters $t$) \cite{Soldini2023}. In fig. S10 (see supplementary text), we also explored disorders or alternative boundaries that can break the mirror symmetry of the system. Since $a$ depends on the molecular spacing, future work could explore the design of the precursor's side groups to access variable sizes and lattice symmetries on alternative superconducting platforms \cite{Liu2023}. Importantly, our findings demonstrate the reversible control of the charge (spin) in radical molecules by the local electric field of the tip, opening interesting avenues for the fine-tuning of the system with external gate voltages. Creating local charge defects in the molecular assembly using probe chemistry \cite{Kawai2020} might also allow to investigate the effect of disorder on the topological phase as well as the topological edge modes. Overall, our work constitutes key advances in designing gate-tunable organic topological superconductors by the self-assembly of organic metal-free molecules in proximity to a superconducting substrate.
 


\section*{Acknowledgments}
{\bf Funding:} E.M. and R.P. acknowledge funding from the Swiss Nanoscience Institute (SNI) and the European Research Council (ERC) under the European Union’s Horizon 2020 research and innovation programme (ULTRADISS grant agreement No 834402 and supports as a part of NCCR SPIN, a National Centre of Competence (or Excellence) in Research, funded by the SNF (grant number 51NF40-180604). E.M., T.G. and S.-X.L. acknowledge the Sinergia Project funded by the SNF (CRSII5\_213533). E.M., T.G. and R.P. acknowledge the SNF grant (200020\_188445). T.G. acknowledges the FET-Open program (Q-AFM grant agreement No 828966) of the European Commission. S.-X. L. acknowledges the grant from the SNF (200021\_204053). J.-C.L. acknowledges funding from the European Union’s Horizon 2020 research and innovation programme under the Marie Sklodowska-Curie grant agreement number 847471. U.A. acknowledges funding by the SNF Professorship (Grant No. PP00P2 187185/2). R.H. acknowledges the European Union’s Horizon 2020 research and innovation programme under Grant Agreement No. 862046 and the ERC grant under Grant Agreement No.757725. Calculations were performed on UBELIX (http://www.id.unibe.ch/hpc), the HPC cluster at the University of Bern. C.L. acknowledges the Georg H. Endress Foundation for financial support. W. W. gratefully acknowledge financial support from the Deutsche Forschungsgemeinschaft (DFG, German Research Foundation) through the Collaborative Research Centre "4f for Future" (CRC 1573, project number 471424360) project B2.  R.P. acknowledges Ruslan Temirov for the fruitful discussions on the improvement of the spectral resolution of the Basel microscope.\\

\noindent    
{\bf Authors contributions:}
R.P., S.-X.L., S.D and E.M. conceived the experiments. P.Z., Ro.H., S.-X.L. and S.D. synthesized the materials. C.D., C.L., and R.P. performed the STM/AFM measurements at 1 K. J.-C.L., H.C. and W.W. performed the millikelvin measurements. U.A. performed DFT calculations. Ri.H., D.L. and J.K. performed tight-binding calculations. R.P. analyzed the data and wrote the manuscript. All authors discussed on the results and revised the manuscript.\\

\noindent    
{\bf Competing interests:}
The authors declare no competing financial interests.\\

\noindent    
{\bf Data and materials availability:}
Data in formats other than those presented within this paper are available from the corresponding authors upon reasonable request. 
\section*{Supplementary materials}
Materials and Methods\\
Supplementary Text\\
Figs. S1 to S15\\
References \textit{(50-69+)}

\bibliography{reference}

\bibliographystyle{Science}


\clearpage

\newpage

\newpage

\newpage

\newpage



\end{document}